\documentclass[12pt]{article}
\usepackage{graphicx}
\usepackage{psfrag}
\usepackage{epstopdf}
\usepackage{latexsym}
\usepackage{slashed}
\usepackage{amsmath}

\begin{document}

\begin{titlepage}
\null\vspace{-62pt}

\pagestyle{empty}
\begin{center}

\vspace{1.0truein} {\Large\bf A comment on lepton mixing}

\vspace{1in}
{\large Dimitrios Metaxas} \\
\vskip .4in
{\it Department of Physics,\\
National Technical University of Athens,\\
Zografou Campus, GR 15773 Athens, Greece\\
 metaxas@central.ntua.gr}\\

\vspace{.5in}
\centerline{\bf Abstract}

\baselineskip 18pt
\end{center}

Since right-handed neutrinos, if added to the Standard Model, have no gauge interactions, their kinetic terms can be mixed. I examine the related rotations of the gauge eigenstates  in order to derive the propagators for the kinetic and mass terms, and I comment on the resulting lepton mixing, on the possibility of not having well-defined mass eigenstates, and on the modifications to weak contributions and observables (anomalous magnetic moment, muon decay, neutrino oscillations).

\end{titlepage}
\newpage
\pagestyle{plain}
\setcounter{page}{1}
\newpage

\section{Introduction}

The wealth and interest, as well as the history of neutrino physics, have produced numerous models and investigations, ranging from smaller to greater extensions of the Standard Model (SM), to applications in cosmology and astrophysics \cite{neutrinos}. 

The minimal extension of the Standard Model includes a right-handed neutrino singlet for each generation, enabling the appearance of a Dirac mass term after symmetry breaking, similar to the process of  mass generation for the other SM fermions. There is a mixing matrix as a result of the rotation between gauge and mass eigenstates  for the leptons (the PMNS matrix) similar to the CKM matrix describing the quark mixing, with interesting physical consequences (CP violation in the lepton sector, for example).

However, there is one important difference for the right-handed neutrino singlets, namely the fact that consistency with the other particle content, charges and anomalies, requires that they have no gauge interactions. The principle of gauge coupling universality is meaningless as far as right-handed neutrinos are concerned; hence the possibility of the  mixing of their kinetic terms will be examined in this work. Another possibility, the addition of Majorana mass terms, and the so-called see-saw mechanism, will not be considered here at first, but will be discussed briefly at the end.

A change of basis, a ``rotation" in particle space, such as the rotation from gauge to mass eigenstates, is generally described by a unitary matrix. Here, with the presence of a general kinetic mixing, a change to mass eigenstates may not be possible, and mixed propagators naturally appear for the neutrinos. This modifies the weak contributions to observables and processes, such as the muon anomalous magnetic moment \cite{amm}, which was the main motivation for this investigation, but also the lepton anomalies \cite{leptonanomaly}, the $\mu\rightarrow e\gamma$ decay, as well as the probabilities for neutrino oscillations \cite{oscillations}.
It may not be possible to satisfy the constraints of all the above-mentioned contributions, even with fine-tuning for both the Yukawa and the kinetic terms; the possibility of fermions not having well-defined mass eigenstates, however, and generally the mixing via the kinetic terms, is of interest by itself and commented upon.

In Sec.~2, I describe the suggested procedure, the derivation of the propagators, and the resulting modifications to the weak calculations. In Sec.~3, I discuss the possible applications and extensions.

\section{Kinetic mixing terms}

The quark sector of the SM consists of the left-handed $SU(2)_L$-doublets 
$
Q_a=
\begin{pmatrix}
 u_{La} \\ 
d_{La}
\end{pmatrix}
$
and the right-handed quark singlets $u_{Ra}, d_{Ra}$, with hypercharges $Y_Q=1/6,\, Y_u=2/3,\, Y_d=-1/3,$ respectively for the $U(1)_Y$ gauge group.
Generally, indices $a, b$ with values $1, 2, 3$ (or $e, \mu, \tau$) will label the three generations and the gauge particle eigenstates.
The Higgs doublet, $H$, with hypercharge $1/2$ and vacuum expectation value 
$
\begin{pmatrix}
0 \\
v
\end{pmatrix}
$
is used, 
together with $\tilde{H}=i \sigma^2 \, H^{*}$, with hypercharge $-1/2$.
The fermion kinetic terms and gauge interactions are written as
\begin{eqnarray*}
\bar{Q}_a(i\slashed{\partial}+g \slashed{W} +g' Y_Q \slashed{B}) Q_a +\\
+\bar{u}_{Ra}(i\slashed{\partial}+ g' Y_u \slashed{B})u_{Ra}+\\
+\bar{d}_{Ra}(i\slashed{\partial}+ g' Y_d \slashed{B})d_{Ra}
\end{eqnarray*}
with the gauge fields $W_{\mu}=W_{\mu}^p \frac{\sigma^p}{2}$ and $B_\mu$
($p=1,2,3$ are the $SU(2)$ indices, $\sigma^{p}$ are the usual Pauli matrices, and $\slashed{a}\equiv a_\mu\,\gamma^\mu$
denotes the product of any vector with the Dirac matrices).

Then, the gauge couplings $g, g'$,
and the Yukawa interactions
\begin{equation}
{Y}^{ab}_d  \bar{Q}_a \, H\, d_{Rb} + Y^{ab}_u \bar{Q}_a \, \tilde{H} u_{Rb} + h.c.
\end{equation}
with the dimensionless coupling constants ($Y_{u,d}$)
after symmetry breaking give the non-diagonal mass terms 
\begin{equation}
\bar{d}_L \, Y_d \,d_R + \bar{u}_L \, Y_u \, u_R + h.c.
\end{equation}

Here, the generation indices $a,b$ have been supressed and the constant $v$ is included in the matrices $Y_{u,d}$, which are 
non-diagonal mass terms since the fermions are still in the gauge eigenstates.
In order to go to the mass eigenstates, the matrices $Y_{u,d}$ are diagonalised by unitary matrices
\begin{equation}
Y_d= U_d M_d K_d^{\dagger}\,, \,\,\, Y_u=U_u M_u K_u^{\dagger}
\label{diag1}
\end{equation}
and the fermion gauge eigenstates are rotated to mass eigenstates by changes of bases
\begin{eqnarray*}
d_{La}&=&U_{d(ai)} \, d_{Li} \\
u_L &=& U_u \, u_L \\
d_R &=& K_d \, d_R \\
u_R &=& K_u \, u_R.
\end{eqnarray*}
Here, in the first change of basis, the indices are written explicitly; $a,b =1,2,3$ (or $e, \mu, \tau$) are used for the gauge eigenstates
and $i,j =1,2,3$ for the mass eigenstates, and they are summed over (whereas the subscripts $u, d$ in (\ref{diag1}) obviously are not).

Then, the mass terms 
\begin{equation}
m_{d,i}(\bar{d}_{Li} d_{Ri} + \bar{d}_{Ri} d_{Ri}) + m_{u,i}(\bar{u}_{Li} u_{Ri} + \bar{u}_{Ri} u_{Li})
\end{equation}
appear, with $m_{d,i}, m_{u,i}$ the diagonal, non-zero, elements of $M_d, M_u$, and the fermion propagators are derived, 
since the left- and right- handed parts can be combined into  Dirac spinors, for example, $u=u_L + u_R$, and together with the kinetic terms,
$\bar{u}(i\slashed{\partial} - m_u) u$ is inverted to give the propagator (with momentum $p$)
\begin{equation}
\frac{i (\slashed{p} + m_u)}{p^2 - m_u^2}
\end{equation}

The rotations affect the interactions of the $u, d$ with the $W^{\pm}$ bosons, and introduce the unitary CKM mixing matrix
\begin{equation}
V_{CKM (ij)} = U^{\dagger}_{u (ia)} \, U_{d(aj)}
\end{equation}
(the other interactions are not affected since they do not mix up and down type quarks).
There are three independent angles in the CKM matrix and generally there are six phases. However, since 
the up and down type quarks can be rotated with  phase factors, except for an overall phase, there is 
finally one independent phase in the CKM matrix.

The mass generation for the lepton sector, with the left-handed $SU(2)_L$-doublets 
$
L_a=
\begin{pmatrix}
 \nu_{La} \\ 
e_{La}
\end{pmatrix}
$
and the right-handed singlets $e_{Ra}, \nu_{Ra}$, proceeds in a similar manner
when the kinetic and gauge terms are
\begin{eqnarray}
\bar{L}_a(i\slashed{\partial}+g \slashed{W} +g' Y_L \slashed{B}) L_a + \notag \\
+\bar{e}_{Ra}(i\slashed{\partial}+ g' Y_e \slashed{B})e_{Ra}+ \notag \\
+\bar{\nu}_{Ra}(i\slashed{\partial}+ g' Y_{\nu} \slashed{B})\nu_{Ra}
\label{sml}
\end{eqnarray}
with the hypercharges $Y_L=-1/2, Y_e = -1, Y_\nu =0$.

The last assignment, $Y_\nu=0$, is a consequence of the fact that neutrinos have zero electric charge, and is also required for the cancellation of the gauge anomalies in the SM.
The Yukawa interactions for the leptons, compatible with their quantum numbers,
\begin{equation}
{Y}^{ab}_e  \bar{L}_a \, H\, e_{Rb} + Y^{ab}_\nu \bar{L}_a \, \tilde{H} \nu_{Rb} + h.c.,
\end{equation}
give Dirac masses to both the charged leptons and neutrinos, once the non-diagonal mass terms 
\begin{equation}
\bar{e}_L \, Y_e \,e_R + \bar{\nu}_L \, Y_\nu \, \nu_R + h.c.
\end{equation}
that appear after symmetry breaking (the $Y$s include a factor of $v$) are diagonalised
by unitary matrices
\begin{equation}
Y_e= U_e M_e K_e^{\dagger}\,, \,\,\, Y_\nu=U_\nu M_\nu K_\nu^{\dagger}
\label{diag2}
\end{equation}
and the fermion gauge eigenstates are rotated to mass eigenstates by changes of bases
\begin{eqnarray*}
e_{La}&=&U_{e(ai)} \, e_{Li} \\
\nu_L &=& U_\nu \, \nu_L \\
e_R &=& K_e \, e_R \\
\nu_R &=& K_\nu \, \nu_R.
\end{eqnarray*}
The lepton terms become
\begin{equation}
\bar{e}_i (i\slashed{\partial} - m_{e,i}) e_i \,+\, \bar{\nu}_i (i \slashed{\partial} - m_{\nu, i}) \nu_i
\end{equation}
with
\begin{equation}
e=e_R + e_L, \,\,\, \nu=\nu_R +\nu_L
\end{equation}
The procedure is entirely analogous to the one in the quark sector and the unitary PMNS mixing matrix appears
\begin{equation}
V_{PMNS (ij)} = U^{\dagger}_{\nu (ia)} \, U_{e(aj)}
\end{equation}
in the $W^{\pm}$ interactions between the left-handed electrons and neutrinos,
parametrised with three independent angles and a phase.

The smallness of the Dirac neutrino masses compared to other scales of the SM, as well as the fact that $Y_\nu =0$, as mentioned before,
sometimes motivates the introduction of neutrino Majorana mass terms, with various mass scales.
They involve the conjugate neutrino fields, and are not allowed if neutrinos carry a quantum number (such as lepton number).

Here, I will address another possibility that arises from the fact that $Y_\nu=0$, namely the 
introduction of dimensionless kinetic mixing terms, $Z^{ab}$, to write the last term in (\ref{sml}) as
\begin{equation}
Z^{ab} \,\bar{\nu}_{Ra}\, i \slashed{\partial} \, \nu_{Rb}.
\label{z1}
\end{equation}
Majorana mass terms will not be included here at first, but commented upon later.

The neutrino kinetic and Yukawa terms become
\begin{equation}
\bar{\nu}_L \, I\, i\slashed{\partial} \nu_L + \bar{\nu}_R \, Z\, i\slashed{\partial} \nu_R + \bar{\nu}_L Y_\nu \,\nu_R + h.c.
\end{equation}
with the identity $3\times 3$ generation matrix written explicitly.
$Z$ is a Hermitian, positive definite matrix, because of positivity constraints, hence it can be diagonalized by a change of basis
\begin{equation}
\nu_{Ra} = K'_{\nu (ai)} \,\nu_{R i},
\label{r1}
\end{equation}
with a unitary matrix $K'_\nu$, to a diagonal matrix, $Z_d$, with positive diagonal entries, not necessarily equal to unity.
For the resulting Yukawa matrix, $Y_\nu \, K'_\nu$, one can use the remaining freedom to rotate the left-handed fields,
after writing it as the product of a unitary times a Hermitian matrix, $Y_\nu \, K'_\nu = U'_\nu \, H_\nu$,
\begin{equation}
\nu_{La} = U'_{\nu (ai)} \, \nu_{L i},
\label{r2}
\end{equation}
in order to get the final expression for the neutrinos
\begin{equation}
\bar{\nu}_L \, I\, i\slashed{\partial} \nu_L + \bar{\nu}_R \, Z_d\, i\slashed{\partial} \nu_R + \bar{\nu}_L H_\nu \nu_R + h.c.
\label{z2}
\end{equation}
There are no additional changes of bases that can simplify the kinetic and Yukawa terms, and the left- and right- handed components of the neutrinos generally do not combine
into a Dirac spinor. The full expression can be inverted, however, in order to get the propagators.
Also, the diagonalization of the electron terms proceeds as usual, and the $W^{\pm}$ interactions with the left-handed electrons and neutrinos still
include a unitary mixing matrix
\begin{equation}
V'_{PMNS (ij)} = U'^{\dagger}_{\nu (ia)} \, U_{e(aj)}.
\label{newpmns}
\end{equation}
The indices $i, j$ here and in (\ref{r1}), also implied in (\ref{z2}),  now do not denote mass eigenstates; they are simply the labels for the new bases.
Indices $a, b$, of course, will still denote the original gauge eigenstates.
The unitary PMNS matrix now still has three independent angles. However, since the three neutrino generations cannot be rotated with independent phase factors,
there are three overall independent phases in the PMNS matrix (this is similar to the situation with Majorana neutrinos).

In order to get a simpler expression for the resulting propagators, I will write the result for the case of two generations,
when the kinetic and mass terms to be inverted  can be written as
\begin{equation}
\begin{pmatrix}
i\slashed{\partial}\, I & H_\nu \\
H_\nu & i \slashed{\partial} \, Z_d,
\end{pmatrix}
\label{twogen}
\end{equation}
with $I$ the $2\times 2$ identity matrix, $Z_d$ the diagonal matrix with elements $Z_1, Z_2$, and 
$
H_\nu =
\begin{pmatrix}
m & x \\
x & M
\end{pmatrix},
$
with real elements with dimensions of mass.
Inverting this expression gives for the upper left $2\times2$ matrix, corresponding
to the left-handed neutrino propagators,
the values
\begin{equation}
G_{11}= \frac{ i\slashed{p}\, (Z_1 Z_2 \,p^2 - Z_1\, M^2 - Z_2 \,x^2)}{D} 
\label{new1}
\end{equation}
\begin{equation}
G_{12} = G_{21} = \frac{ i \slashed{p} \, x\, (Z_1 \,M + Z_2\,m)}{ D}
\label{new2}
\end{equation}
\begin{equation}
G_{22}= \frac{ i\slashed{p}\, (Z_1 Z_2 \,p^2 - Z_1\, x^2 - Z_2 \,m^2)}{D} 
\label{new3}
\end{equation}
with
\begin{equation}
D=Z_1\,Z_2\, p^4 - p^2 \left((Z_1 +Z_2) x^2 +Z_1 \,M^2 +Z_2\, m^2 \right)
                           + (m\,M - x^2)^2
\label{d}
\end{equation}
Generally, mixed neutrino propagators, $G_{ij}$, appear and the relevant electroweak processes are modified.
The propagators derived here are to be compared with the usual Dirac neutrino propagators
\begin{equation}
G_i =\frac{i (\slashed{p} + m_i)}{p^2 - m_i^2},
\label{usual}
\end{equation}
where the $i$'s label mass eigenstates with well-defined neutrino mass, $m_i$.
The usual contribution for the muon anomalous magnetic moment, for example, includes the factors
\begin{equation}
V^{\dagger}_{\mu i} \, G_i \, V_{i \mu},
\end{equation}
 related to the internal neutrino line,
whereas the modified contribution becomes
\begin{equation}
V'^{\dagger}_{\mu i} \, G_{ij} \, V'_{j \mu}.
\label{newew}
\end{equation}
Similar changes appear in the other electroweak processes, as in Fig.~1, where the solid lines denote
leptons $e, \mu, \tau$, wiggly lines denote $W^{\pm}$'s of photons, and double solid lines denote neutrinos.
The mixing matrices $V, V'$, in any case, remain unitary, of course, as explained before.
The possibility of enhancing the electroweak contribution to the muon anomalous magnetic moment as above, should be 
considered together with the constraints from other processes, such as the $\mu\rightarrow e\gamma$, for  example, where the usual contribution of the similar diagram is suppressed with
\begin{equation}
V^{\dagger}_{\mu i} \, G_i \, V_{i e},
\end{equation}
whereas the modified contribution becomes
\begin{equation}
V'^{\dagger}_{\mu i} \, G_{ij} \, V'_{j e}.
\end{equation}
The mixed propagators with the common denominator (\ref{d}) may have negative metric or ghost-like poles or residues for some values of the parameters. It should be remembered, though, that they always appear in combinations and sums for various internal lines, and the consistency of the entire final expressions for the amplitudes should be checked.

\section{Comments and order of magnitude estimates}

As far as the modifications in the electroweak contributions to the muon anomalous magnetic moment are concerned, one can make some preliminary, order of magnitude estimates, of the previous considerations, in some limiting cases.

First, one may consider the case where $Z_1, Z_2 =1$ and 
$x^2 >> M^2, m^2$ with both $M^2, m^2 \sim \mu^2$
in equations (\ref{twogen}, \ref{new1}, \ref{new2}, \ref{new3}) and (\ref{d}).
Here, $\mu$ is a scale related to the neutrino mass, although the latter, as discussed before and in the next Section,
is not well-defined, and neutrinos are always mixed.

In order to compare with the usual, one-loop, electroweak contribution, involving the $W$-boson, to the muon anomalous magnetic moment,
\begin{equation}
a_W=\frac{10}{3}\frac{g^2 m_\mu^2}{64 \pi^2 m_W^2}
\end{equation}
($m_\mu$ and $m_W$ are, respectively, the muon and $W$-boson masses, in the first diagram of Fig.~1)
one may also consider the limit where
$M_W >> x >> \mu$
where the propagators simplify to $G_{11}=G_{22}=\frac{i \slashed{p}}{p^2 - x^2}$,
$G_{12}=G_{21}=\frac{i \slashed{p} x \mu}{(p^2 - x^2)^2}$.

The simplifying case of two generations is also still assumed, with a $2\times 2$ unitary mixing matrix (\ref{newpmns})
involving an angle $\theta$ (the analogue of an electroweak ``Cabibbo angle'').
Then, the diagonal elements in (\ref{newew}) reproduce the standard electroweak result, $a_W$,
and the off-diagonal terms give an additional contribution of the order of
$\frac{x \mu}{M_W^2} \sin\theta \cos \theta \, a_W$.

As another limiting case, one may also consider $Z_1=1$ and $Z_2=0$, again with 
$M_W >> x >> \mu$, when the propagators become
$G_{11}=\frac{i \slashed{p} \mu^2}{x^2 (p^2 -x^2)}$,
$G_{22}=\frac{i \slashed{p}}{p^2-x^2}$ and $G_{12}=G_{21}=\frac{i \slashed{p} \mu}{x(p^2- x^2)}$.
Then the diagonal contribution to the muon anomalous magnetic moment
becomes
$(\sin^2\theta \frac{\mu^2}{x^2} + \cos^2 \theta) a_W$,
and the off-diagonal $\frac{2\mu}{x} \sin\theta \cos\theta \,a_W$.

In both limits considered here, the modifications are subleading compared to the usual, one-loop, electroweak result (in fact, the second limit also modifies the standard result, approximately by a factor of $\cos^2 \theta$). It should be noted, however, that, although it seems natural to assume that $M_W$ is much larger than $x, \mu$, the role of the latter parameters and their relation and interpretation as neutrino masses is still not clear, as well as their order of magnitude compared to other parameters of the electroweak fermion sector. A more complete investigation is necessary, considering other processes involving neutrinos and neutrino oscillations and their contributions in lepton mixings and violations of lepton universality in the physics of the Standard Model, phenomenological models and unified theories that extend it or include it as a low-energy effective theory.
Finally, it should also be noted that the proposal of a general kinetic mixing term suggested here may arise from decompactifications of higher-dimensional string-inspired theories, brane-antibrane effects, or even as an effective theory stemming from similar considerations \cite{strings}.

\section{Discussion}

The concept of particle masses and mass scales, their generation and definition through the action or effective action of a theory is central in particle physics \cite{mass}.
In this work, I presented a modification of the Standard Model Lagrangian, including right-handed neutrino singlets
with mixed kinetic terms. This leads to generalized, mixed propagators for the neutrinos, and the absence of well-defined 
mass eigenstates, although a unitary mixing matrix, corresponding to the related changes of bases, is still involved.
One may conjecture that right-handed singlets have different ``normalization" than other SM particles, originating from some process of dimensional or other kind of reduction from a more complete theory.

Besides the theoretical search for possible interpretations of such mixings, there is also the question of the experimental results where one does not always observe or measure an entire process involving one or more Feynman diagrams, with loops and intermediate particles, but rather an initial or final external state; apart from problems of coherence, do neutrinos behave as Weyl particles there, having $p^2 =0$, or do they ``pick" one of the pole mass terms?  The treatment given here suggests that neutrinos are massless particles in external states, and behave as massive particles in intermediate states, although without a well-defined mass term but a combination (or superposition) thereof. This still leaves several questions open regarding their thermal and statistical behavior, their contributions in cosmological settings, relic abundancies, dark matter, etc.

The main motivation here was to investigate possible discrepancies and anomalies between theoretical and experimental values of electroweak observables,
such as the muon anomalous magnetic moment. For example, the magnitude of the mass terms in the internal neutrino states may be arbitrary and contribute to these quantities, although the external states still remain massless. However, the agreement with already existing results in other 
processes, such as neutrino oscillations or the $\mu\rightarrow e \gamma$ decay, may still not be possible, or require additional fine-tuning of the parameters involved.

In particular, as far as neutrino oscillations are concerned, they happen regardless of the relative magnitudes of the mass parameters involved, and the related probabilities should be calculated using the field-theoretical approach \cite{fto}, although quantum-mechanical questions, like the ones mentioned above, are pertinent.

I have not considered the further inclusion of Majorana neutrino mass terms, which is still possible. The usual addition of these terms involves additional mass scales, much larger than the SM  scales, which lead to much smaller neutrino masses, by the so-called
see-saw mechanisms. Then, there appears a small deviation from unitarity for the leptonic mixing matrices, suppressed by the
same ratio of widely different scales. If one considers, furthermore, the mixing of the kinetic terms, as described here, it is expected that possibly larger deviations from unitarity will be present, giving similar corrections to electroweak observables, and demanding again a comparison between the theoretical and experimental constraints.  Majorana mass eigenstates are still defined, however, since the neutrinos can now be rotated together with their antiparticles, being indistinguishable from them. The full $6\times 6$ matrix for the change of basis is still unitary, but the upper left $3\times 3$ submatrix deviates from unitarity, giving again modified  contributions.

These and other investigations will hopefully be the subject of future work; I would like in closing to mention again the fact that
the apparent lack of well-defined neutrino mass eigenstates that  combine the left- and right- handed spinors is of theoretical interest on its own, regardless of possible phenomenological applications in SM physics. In particular, one may further investigate questions and problems regarding the different nature and properties of Weyl, Dirac, and Majorana spinors, their possible combinations, their behavior in processes involving loop diagrams, tree diagrams, internal or external states, etc.

\newpage

\begin{figure}
\centering
\includegraphics[width=60mm]{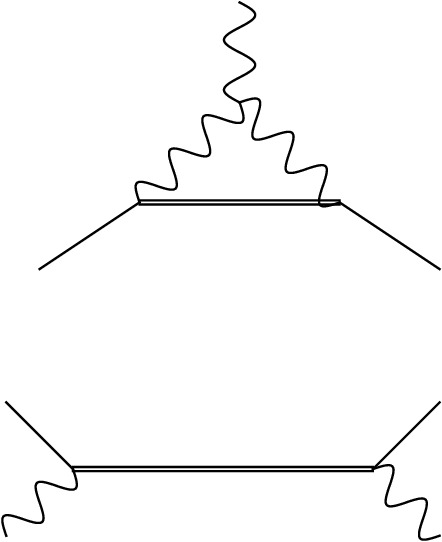}

\caption{   The diagrams involved in the calculation of the anomalous magnetic moment or the rate of $\mu \rightarrow e\gamma$ decay, and in neutrino oscillations. Wiggly lines denote $W^{\pm}$'s and an external photon in the first diagram, solid lines denote leptons $e, \mu, \tau$, and double solid lines neutrinos. }
\end{figure}


\vspace{0.5in}




\begin{thebibliography}{99}

\bibitem{neutrinos} R.~N.~Mohapatra {\it et al.}, {\it Rept. Prog. Phys.} {\bf 70}, 1757 (2007).

                             K.~N.~Abazajian {\it et al.}, e-Print: 1204.5379 [hep-ph].
    
                            M.~Dvornikov and A.~Studenikin, {\it Phys. Rev.} {\bf D69}, 073001 (2004).
   
                            H.~Swami, K.~Lochan and K.~M.~Patel, {\it Phys. Rev.} {\bf D102}, 024043 (2020).
    
                           W.~L.~Xu, J.~B.~Mu\~noz and C.~Dvorkin, {\it Phys.Rev.} {\bf D105}, 095029 (2022).
  
                          T.~Brinckmann, J.~H.~Chang and M.~LoVerde, {\it Phys. Rev.} {\bf D104}, 063523 (2021).
    
                                          
\bibitem{amm}     F.~Jegerlehner and A.~Nyffeler, {\it Phys. Rept.} {\bf 477}, 1 (2009).

                           T.~Aoyama {\it et al.}, {\it Phys. Rept.} {\bf 887}, 1 (2020).
   
                           Y.~Farzan and M.~Tortola, {\it Front. in Phys.} {\bf 6}, 10 (2018).
  
                           C.~H.~Chen and T.~Nomura, {\it Nucl. Phys.} {\bf B964}, 115314 (2021).
                            
                           L.~Darm\'e, M.~Fedele, K.~Kowalska and E.~M.~Sessolo, {\it JHEP} {\bf 03},085 (2022).
   
                           L.~T.~ Hue, K.~H.~Phan, T.~P.~Nguyen, H.~N.~Long and H.~T.~Hung, {\it Eur.Phys.J.} {\bf C82}, 722 (2022).
  
                          M.~Davier, A.~Hoecker, B.~Malaescu and Z.~Zhang, {\it Eur. Phys. J.} {\bf C77}, 827 (2017).

                         
                          

\bibitem{leptonanomaly}              R.~Aaij {\it et al.} (LHCb), {\it Phys. Rev. Lett.} {\bf 122}, 191801 (2019).
   

                                      S.~Descotes-Genon, S.~Fajfer, J.~F.~Kamenik and M.~Novoa-Brunet, {\it Phys. Lett.} {\bf B809}, 
                                         135769 (2020).          

                                            M.~ Jung and D.~M.~Straub, {\it JHEP} {\bf 01}, 009 (2019).
   

                                     D.~ Bhatia, N.~Desai and A.~Dighe, {\it JHEP} {\bf 04}, 163 (2022).
   

                                      L.~Allwicher, P.~Arnan, D.~Barducci and M.~Nardecchia, {\it JHEP} {\bf 10}, 129 (2021).

                                    B.~A.~Kniehl and A.~Pilaftsis, {\it Nucl. Phys.} {\bf B474}, 286 (1986).
   

\bibitem{oscillations}       S.~M.~ Bilenky and B.~Pontecorvo, {\it Phys. Rept.} {\bf 41}, 225 (1978).
                                      
                                     M.~Dvornikov, {\it Phys. Rev.} {\bf D100}, 096014 (2019).
    
                                  G.~Koutsoumbas and D.~Metaxas, {\it Gen. Rel. Grav.} {\bf 52}, 102 (2020).
    
                                     H.~Chakrabarty, D.~Borah,  A.~Abdujabbarov, D.~Malafarina and B.~Ahmedov, {\it Eur.Phys.J.} {\bf C82}, 24 (2022).
             
 
                                  M.~Dvornikov, {\it Phys. Rev.} {\bf D104}, 043018 (2021).

\bibitem{strings}        K.~R.~Dienes, C.~F.~Kolda and J.~March-Russell, {\it Nucl. Phys.} {\bf B492}, 104 (1997).
    
                                T.~G.~Rizzo, {\it Phys. Rev.} {\bf D59}, 015020 (1998).

                                S.~A.~Abel and B.~W.~Schofield, {\it Nucl. Phys.} {\bf B685}, 150 (2004).

                               S.~A.~Abel, M.~D.~Goodsell, J.~Jaeckel, V.~V.~Khoze and A.~Ringwald, {\it JHEP} {\bf 07}, 124 (2008).

                                 T.~G.~Rizzo, {\it Phys. Rev.} {\bf D106}, 035024 (2022).

\bibitem{mass}          G.~F.~Giudice, {\it PoS LHCP2021}, 019 (2021) e-Print: 2109.07176 [hep-ph].
                                  
                                 S.~R.~Coleman and E.~J.~Weinberg, {\it Phys. Rev.} {\bf D7}, 1888 (1973).

                                 P.~Gambino and P.~A.~Grassi, {\it Phys. Rev.} {\bf D62}, 076002 (2000).

                                 D.~Metaxas, {\it Phys. Lett.} {\bf B816}, 136243 (2021).

                                 D.~Metaxas, {\it Phys. Rev.} {\bf D98}, 036001 (2018).
  
                                 H.~Shen, Y.~Cheng and W.~Liao, {\it Phys. Rev.} {\bf D103}, 076016 (2021).
   
  
\bibitem{fto}              M.~Beuthe, {\it Phys. Rept.} {\bf 375}, 105 (2003).
   
                                  C.~Giunti, {\it JHEP} {\bf 11}, 017 (2002).
    
                                 M.~Blasone, A.~Capolupo and G.~Vitiello, {\it Phys. Rev.} {\bf D66}, 025033 (2002).
  
\end{thebibliography}
\end{document}